\documentclass[12pt]{iopart}
\begin{document}

\title[Classification of entanglement and quantum phase transition in
$XX$ model]{Classification of entanglement and quantum phase
transition in $XX$ Model}

\author{Ting Zhang, Pinx-Xing Chen, Wei-Tao Liu and Cheng-Zu Li}
\address{\small\emph{Department of Physics, Science college,
National University of Defence Technology, Changsha 410073, China}}
\ead{ccat1980zt@gmail.com}

\begin{abstract} We study the relation
between entanglement and quantum phase transition (QPT) from a new
perspective. Motivated by one's intuition: QPT is characterized by
the change of the ground-state structure, while entangled states
belong to different classes have different structures, we conjecture
that QPT occurs as the class of ground-state entanglement changes
and prove it in $XX$ model. Despite the classification of
multipartite entanglement is yet unresolved, we proposed a new
method to judge whether two many-body states belong to the same
entanglement class.
\end{abstract}


\maketitle

\section{Introduction}

Quantum phase transition (QPT) is a phase transition that occurs at
absolute zero temperature, and means nonanalyticity of the
ground-state properties. The singularity may be a discontinuity in
the first or higher-order derivative of the ground-state energy,
respectively referred as first-order or continuous QPT\cite{QPT}.

QPT is usually accompanied by a qualitative change in the nature of
the correlation in the ground state, and quantum fluctuation is the
ultimate reason leading to QPT, so it is certainly of major interest
in both condensed matter physics and quantum information science to
describe the connection between QPT and quantum entanglement
\cite{book}. Various entanglement measures are calculated and hoped
to exhibit singular behavior at quantum critical point. Some
bipartite entanglement measurements such as
concurrence\cite{concurrence1,concurrence2,concurrence3},
entanglement entropy\cite{entanglement entropy1,entanglement
entropy2}, can indeed identify particular QPT. There even a general
theory of the relation between QPT and bipartite entanglement was
developed under certain conditions\cite{PRL_93_250404}. However,
counterexample was soon found\cite{PRA_71_030302}. Therefore, QPT in
terms of multipartite entanglement began drawing
attentions\cite{mulitipartite entanglement1,mulitipartite
entanglement2}. It is also a noteworthy problem whether multipartite
or bipartite entanglement being favored at QPT\cite{B or M1,B or
M2}.

So far there is not a universal conclusion of the relation between
entanglement (bipartite or multipartite) and QPT (first-order or
continuous). Considering that the intrinsic character of QPT is the
change of the ground-state structure, so the problem is probably due
to that the singularities of entanglement measures aren't essential
to the change of the ground-state structure. Geometric
phase\cite{GP} or fidelity\cite{fidelity1,fidelity2,fidelity3} may
be a good indication of QPT. But concerning entanglement, we think
that inequivalent entanglement class can appropriately reflect the
change of ground-state structure, since two inequivalently entangled
states have different structures. So a promising way to reveal the
deep connection between QPT and entanglement is to study the
classification of ground-state entanglement around QPT. Our focus in
the paper is to investigate the classification of entangled ground
states in the vicinity of QPT in one-dimensional $XX$ model
\cite{XY}. It is shown that, no matter the model length $N$ is
arbitrary or tends to infinity, the change of ground-state
entanglement class always indicates the occurrence of QPT.

\section{SLOCC classification of entanglement}
As concerning the classification of entangled states, stochastic
local operation and classical communications (SLOCC) are usually
used to define equivalent classes. That is, two states are said to
belong to the same entanglement class if both of them can be
obtained from the other with nonzero probability by means of local
operation assisted by classical communications. Many researchers
have investigated the SLOCC-inequivalent classes of pure entangled
states \cite{SLOCC1,SLOCC2,SLOCC3}.

The complete classification of pure entangled states is indeed an
intricate task. However, if we only need to judge whether two states
are SLOCC-inequivalent, there is a simple criteria by virtue of
Schmidt decomposition \cite{book}. As well known, for any bipartite
pure state $|\Psi\rangle\in \mathcal{H}=\mathcal{H}_A \otimes
\mathcal{H}_B$, there exist local orthonormal bases $\{|u_i\rangle\}
\in \mathcal{H}_A$ and $\{|v_i\rangle\} \in \mathcal{H}_B$ such that
\begin{equation}\label{1}
|\Psi\rangle=\sum\limits_{i} a_{i} |u_i\rangle \otimes |v_i\rangle,
\end{equation}
$a_i$ are nonnegative real numbers satisfying $\sum_i a_i^2=1$,
referred to as Schmidt coefficients. The number of nonzero $a_i$ is
known as the Schmidt rank, here denoted by Sch$(|\Psi\rangle)$. It
can be easily deduced that any two SLOCC-equivalent states must have
the same Schmidt rank \cite{SSN1,SSN2}. In other words, two
bipartite states with different Schmidt rank are SLOCC-inequivalent.
It provides a clue to judge the SLOCC-inequivalence of multipartite
state. Given two $N$-party states, we can calculate its Schmidt rank
based on a particular bipartition. A bipartition means a division of
$N$-party system into two nonempty and disjoint parts, i.e., one
part including $M (1\leq M<N)$ bodies and the other $(N-M)$ bodies.
If two $N$-party states have different Schmidt rank based on the
same bipartition, they necessarily belong to different entanglement
class. This method bypasses the involved issue of complete
classification of multipartite states and may be crucial in the
researches of QPT in many-body system.

\section{First-order and continuous QPT in $XX$ model }
The Hamiltonian of $XX$ model is
\begin{equation}\label{2}
\mathcal{H}=\frac{J}{4}\sum_{i=1}^N
(\sigma^{i}_{x}\sigma^{i+1}_{x}+\sigma^{i}_{y}\sigma^{i+1}_{y})-B\sum_{i=1}^N
\sigma_{z}^{i}.
\end{equation}
where $\sigma _{x/y/z}^{i}$ are the usual Pauli matrices of the
$i$th spin (cyclic boundary condition $N+1 \equiv 1$ is assumed).
The external magnetic field $B$ could always be supposed positive
without loss of generality. The model can be analytically solved by
Jordan-Wigner transformation \cite{XY}. Using the operators
\begin{eqnarray*}
\sigma_{\pm}&=&\frac{1}{2}(\sigma_{x}\pm \rmi\sigma_{y}),\quad
c_k^{\dag}=\sigma_{-}^{k}\prod\limits_{i=1}^{k-1}\sigma_{z}^{i},
\end{eqnarray*}
the Hamiltonian (\ref{2}) is transformed into
\begin{equation*}
\mathcal{H}=-\frac{J}{2}\sum_{i=1}^{N-1}(c_{i+1}^{\dag}c_i+c_i^{\dag}c_{i+1})+\frac{J}{2}\alpha(c_1^{\dag}c_N+c_N^{\dag}c_1)
-BN+2B\sum_{i=1}^N c_i^{\dag}c_i
\end{equation*}
where
$\alpha\equiv\prod_{k=1}^N(1-2c_k^{\dag}c_k)=\prod_{k=1}^N\sigma_z^k=(-1)^r$.
$r$ is the total number of spin-downs which is a constant.
Introducing the Fourier transformation of $c_k^{\dag}$:
\begin{eqnarray}\label{3}
C_q^{\dag}&=&\frac{1}{\sqrt{N}}\sum\limits_{k=1}^{N}\exp(\rmi
qk)c_k^{\dag},\quad q=\frac{2\pi n}{N}
\end{eqnarray}
where $n$ is integer (half-odd integer) for odd (even) $r$, the
Hamiltonian is that of one-dimensional spinless ferminons
\begin{equation*}
\mathcal{H}=-BN+\sum_q(2B-J\cos q)C_q^{\dag}C_q.
\end{equation*}
The lowest energy eigenstate with fixed $r$ can be expressed as
\begin{equation}\label{4}
|\psi_0^r\rangle=\prod\limits_{l=1}^r
C_{\pi(r+1-2l)/N}^{\dag}|\underbrace{0\cdots 0}_{N}\rangle,
\end{equation}
and its energy is
\begin{eqnarray*}
E_0^r &=&-J\sum_{l=1}^{r}\cos[\pi(r+1-2l)/N]-B(N-2r)\nonumber\\
&=&-J\csc\left(\frac{\pi}{N}\right)\sin\left(\frac{\pi r}{N}\right)-B(N-2r)\nonumber\\
&\hat{=}& -D^{r}J-B(N-2r).
\end{eqnarray*}
Obviously,$D^r=D^{N-r}$. Thus when $J$ is fixed and $B$ is tuned,
\begin{equation}
\mathcal{G} \hat{=} \{|\psi_0^0\rangle,
|\psi_0^1\rangle,\ldots,|\psi_0^{\lfloor\frac{N}{2}\rfloor}\rangle\}\nonumber
\end{equation}
includes all of the possible ground states of the $XX$ model. With
the modulation of the external magnetic field, each state in
$\mathcal{G}$ becomes the ground state in turn, while the energy of
the system changes abruptly. The first-order derivative of the
energy with the magnetic field $B$ is discontinuous. A first-order
QPT occurs.

For $N$-limited $XX$ model, the first-order QPT occur at
$\lfloor\frac{N}{2}\rfloor$ critical values of magnetic field, which
can be achieved from the eigenenergy corresponding to each state in
$\mathcal{G}$. That is, the energy of $|\psi_0^r\rangle$ is
$-D^{r}J-B(N-2r)$, then the transition from $|\psi_0^r\rangle$ to
$|\psi_0^{r+1}\rangle$ occurs at
\begin{equation*}
-D^{r}J+2Br=-D^{r+1}J+2B(r+1),
\end{equation*}
i.e.,
\begin{equation*}
B_c^r=\frac{J}{2}\sec\left(\frac{\pi}{2N}\right)\cos\left[\frac{\pi
(r+\frac{1}{2})}{N}\right],\quad
r=0,\ldots,\lfloor\frac{N}{2}\rfloor-1.
\end{equation*}

For $N\rightarrow\infty$, it is already known that at $B_c=J/2$,
there occurs a continuous QPT, i.e., a superfluid-Mott insulator
phase transition\cite{QPT}.

\section{SLOCC classification of ground-state around QPT in $XX$ model}
Now we begin to investigate the ground-state entanglement class
around QPT in $XX$ model.

When $N\rightarrow\infty$, there is only one contimuous QPT. At
$B_c=J/2$, the system transits from Mott-insulator phase to
superfluid phase or vice versa. The ground state of Mott-insulator
phase is a separable state with all spins pointing to the same
direction, while the ground state of superfluid phase is sure to be
an entangled state\cite{QPT}. Therefore, the class of ground-state
entanglement changes when QPT happens.

When $N$ is limited, since every stats in $\mathcal{G}$ could be
ground state with the adjustment of magnetic field, we need to prove
that all states in $\mathcal{G}$ is inequivalently entangled.

To complete this tough task for arbitrary $N$, we calculate the
Schmidt rank of every element in $\mathcal{G}$, based on
$\lfloor\frac{N}{2}\rfloor\otimes N-\lfloor\frac{N}{2}\rfloor$
bipartition. We will show that Sch$(|\psi_0^r\rangle)=2^r$ holds for
arbitrary $r$ and $N$. For simplicity,
$\lfloor\frac{N}{2}\rfloor\hat{=}M$ henceforth.

Notice that
\begin{eqnarray*}
c_{m}^{\dag}c_{n}^{\dag}|\underbrace{0\cdots 0}_{N}\rangle=\left\{\begin{array}{cc}|mn\rangle,\ \ & \ \ {\rm if}\ \ m<n,\\
-|mn\rangle,\ \ & \ \ {\rm if}\ \ m>n,\\0,\ \ & \ \ {\rm if}\ \
m=n,\\ \end{array}\right.
\end{eqnarray*}
where $|mn\rangle$ corresponds to spin configuration in which all
spins are up, except the spin at the site $m$ and $n$ are down.
Replacing Eq.(\ref{3}) into (\ref{4}), the ground state with $r$
spin-downs can be expressed as
\begin{eqnarray*}
\fl |\psi_0^r\rangle&=&\frac{1}{\sqrt{N^r}}\prod\limits_{l=1}^r
\sum\limits_{k=1}^N\exp\left[\frac{\rmi(r+1-2l)k\pi}{N}\right]c_k^{\dag}|\underbrace{0\cdots
0}_{N}\rangle\nonumber\\
\fl &=&\frac{1}{\sqrt{N^r}}\sum\limits_{1\leq k_1<\ldots<k_r\leq
N}\exp\left[\frac{\rmi(r-1)k_1\pi}{N}\right]\exp\left[\frac{\rmi(r-3)k_2\pi}{N}\right]
\cdots\exp\left[\frac{\rmi(1-r)k_r\pi}{N}\right]|k_1\cdots k_r\rangle\nonumber\\
\fl &=&\frac{(2\rmi)^{C_r^2}}{\sqrt{N^r}}\sum\limits_{1\leq
k_1<\ldots<k_r\leq N}\prod_{1\leq i<j\leq
r}\sin\left[\frac{(k_i-k_j)\pi}{N}\right] |k_1\cdots k_r\rangle.
\end{eqnarray*}

The last step is achieved using $\rme^{\rmi x}-\rme^{-\rmi
x}=2\rmi\sin x$.

Next we calculate the Schmidt rank of $|\psi_0^r\rangle$ based on
$M\otimes N-M$ bipartition. The constant
$(2\rmi)^{C_r^2}/\sqrt{N^r}$ can be omitted. $|\psi_0^r\rangle$ is a
weighted superposition of all possible $|k_1\cdots k_r\rangle$ where
the value range of all $k_i$ are $[1,N]$ and $k_1<\cdots<k_r$ must
be satisfied , so we rewrite $|\psi_0^r\rangle$ as
\begin{eqnarray*}
\fl |\psi_0^r\rangle=|\underbrace{0\cdots
0}_{M}\rangle\otimes\left[\sum\limits_{M<k_1<\ldots<k_r\leq
N}\prod_{1\leq i<j\leq r}\sin
\left[\frac{(k_i-k_j)\pi}{N}\right]|k_1\cdots k_r\rangle\right]\nonumber\\
+\cdots+\sum_{1\leq k_1<\cdots<k_l\leq M}\left[|k_1\cdots
k_l\rangle\otimes\sum_{M<k_{l+1}<\cdots<k_r\leq N}\prod_{1\leq
i<j\leq r}
\sin\left[\frac{(k_i-k_j)\pi}{N}\right]|k_{l+1}\cdots k_r\rangle\right]\nonumber\\
+\cdots+\left[\sum_{1\leq k_1<\cdots<k_r\leq M}\prod_{1\leq i<j\leq
r}\sin\left[\frac{(k_i-k_j)\pi}{N}\right]|k_1\cdots
k_r\rangle\right] \otimes|\underbrace{0\cdots 0}_{N-M}\rangle
\end{eqnarray*}

That is, we first carry out a preliminary Schmidt decomposition by
divide all possible $|k_1\cdots k_r\rangle$ into $r+1$ groups,
according to the number of spin-downs that locate in the former $M$
qubits, i.e.,
\begin{equation}\label{5}
|\psi_0^r\rangle=\sum_{l=0}^r \sum_i a_i^l |u_i^l\rangle \otimes
|v_i^l\rangle,
\end{equation}
Obviously, when $k\neq l$, $\forall i,j$, $\langle
u_i^k|u_j^l\rangle=\langle v_i^k|v_j^l\rangle=0$ always hold. If
Sch$(|\psi_0^{r(l)}\rangle)$ represents the number of nonzero
$a_i^l$, then
\begin{equation}\label{6}
 {\rm Sch}(|\psi_0^r\rangle)=\sum_{l=0}^r {\rm Sch}(|\psi_0^{r(l)}\rangle).
\end{equation}
We will explain that
\begin{equation}\label{7}
{\rm Sch}(|\psi_0^{r(l)}\rangle)=C_r^l,
\end{equation}
thereby
\begin{equation}\label{8}
{\rm Sch}(|\psi_0^{r}\rangle)=\sum_{l=0}^rC_r^l=2^r.
\end{equation}

Notice that the maximum value of $r$ is $M$, so the Schmidt rank of
$|\psi_0^r\rangle$ will never exceed $2^M$.

First it is easily found Sch$(|\psi_0^{r(0)}\rangle)=$
Sch$(|\psi_0^{r(r)}\rangle)=1$. Eq. (\ref{7}) holds for $l=0,r$.

Next remembering that if $\{\alpha_1,\cdots,\alpha_l\}$ is a set of
linear independent vectors, then we can achieve an equivalent set of
orthogonal vectors $\{\beta_1,\cdots,\beta_l\}$ by Schmidt
orthogonalization. Thereby, Sch$(|\psi_0^{r(l)}\rangle)$ is the rank
of such a $C_M^l\otimes C_{N-M}^{r-l}$-dimensional matrix
$A^{r(l)}$. Every element of $A^{r(l)}$ can be uniformly expressed
as $\prod_{1\leq i<j<\leq r}\sin[(k_i-k_j)\pi/N]$, i.e., a product
of $C_r^2$ sine functions. For each row $1\leq k_1<\cdots<k_l\leq M$
are fixed and $M<k_{l+1}<\cdots<k_r\leq N$ vary, while for each
column the situation is just the reverse. We find the rank of
$A^{r(l)}$ by elementary row(column) transformation. So every
element can first be simplified as $\prod_{1\leq i\leq l, l+1\leq
j\leq r}\sin[(k_i-k_j)\pi/N]$, i.e., a product of $l(r-l)$ sine
functions, as for each row of $A^{r(l)}$, $\prod_{1\leq i<j\leq
l}\sin[(k_i-k_j)\pi/N]$ is a constant, and for each column
$\prod_{l+1\leq i<j\leq r}\sin[(k_i-k_j)\pi/N]$ is a constant.
Furthermore,
$\rm{Sch}(|\psi_0^{r(l)}\rangle)=\rm{Sch}(|\psi_0^{r(r-l)}\rangle)$
should hold. The reason is that the elementary row and column
transformation of $A^{r(l)}$ and $A^{r(r-l)}$ will yield similar
simplest form. Then we only need to calculate
$\rm{Sch}(|\psi_0^{r(l)}\rangle)$ by elementary row transformation
for $l=1,\cdots,\lfloor \frac{r}{2}\rfloor$.

Take $r=2,l=1$ as an example, every element can be expressed as
$\sin[(k_1-k_2)\pi/N]$, and for each column $1\leq k_1\leq M$ varies
while for each row $M<k_2\leq N$ varies. So the the matrix is
(overall minus sign is omitted)
\begin{eqnarray*}
A^{2(1)}=\left[\begin{array}{cccc}\sin(\frac{\pi}{N})&\sin(\frac{2\pi}{N})&\cdots&\sin(\frac{N-M}{N}\pi)\\
\sin(\frac{2\pi}{N})&\sin(\frac{3\pi}{N})&\cdots&\sin(\frac{N-M+1}{N}\pi)\\
\vdots&\vdots&\ddots&\vdots\\
\sin(\frac{M}{N}\pi)&\sin(\frac{M+1}{N}\pi)&\cdots&\sin(\frac{N-1}{N}\pi)
\end{array}\right]
\end{eqnarray*}
Let the $i$th row vector of $A^{2(1)}$ is denoted as $\mathbf{a}_i$,
because
\begin{equation}\label{9}
\sin\left(\frac{m\pi}{N}\right)+\sin\left[\frac{(m+2)\pi}{N}\right]=2\cos\left(\frac{\pi}{N}\right)\sin\left[\frac{(m+1)\pi}{N}\right],
\end{equation}
so for $i=1,\ldots M-2$,
\begin{equation}\label{10}
 \mathbf{a}_i+\mathbf{a}_{i+2}=2\cos\left(\frac{\pi}{N}\right)\mathbf{a}_{i+1}
\end{equation}
always holds. Then by elementary row transformation
$\mathbf{a}_{i+2}\hat{=}\mathbf{a}_{i+2}+\mathbf{a}_i-2\cos\left(\frac{\pi}{N}\right)\mathbf{a}_{i+1}$
for $i=1,\cdots,M-2$, $A^{2(1)}$ can be transformed into
\begin{eqnarray*}
A^{2(1)}=\left[\begin{array}{cccc}0&0&\cdots&0\\\vdots&\vdots&\ddots&\vdots\\0&0&\cdots&0\\
\sin(\frac{M-1}{N}\pi)&\sin(\frac{M}{N}\pi)&\cdots&\sin(\frac{N-2}{N}\pi)\\\sin(\frac{M}{N}\pi)&\sin(\frac{M+1}{N}\pi)&\cdots&\sin(\frac{N-1}{N}\pi)
\end{array}\right]
\end{eqnarray*}
So the rank of $A^{2(1)}$ is $2$.  Elementary column transformation
apparently yields the same result.

For bigger $r$ and $l$, although the matrix $A^{r(l)}$ becomes
complicated rapidly, the knack used to find the rank of $A^{r(l)}$
is analogous, however, much more intricate. We will expatiate step
by step.

First consider $A^{r(1)}$, every element is in the simplified form
of $\prod_{2\leq j\leq r}\sin[(k_1-k_j)\pi/N]$, i.e., a product of
$r-1$ sine functions. It can be transformed into a linear sum of a
series of sine functions like
$\sin\{[pk_1+f_p(k2,\cdots,k_r)]\pi/N\}$, where
$p=r-1,r-3,\cdots,1(0)$ for even (odd) $r$. After elementary row
transformation like above, there are two rows left for every nonzero
$p$ and only one row left when $p=0$. Then the simplest expression
of $A^{r(1)}$ is
\begin{eqnarray*}
 A^{r(1)}=\left[\begin{array}{cc}\prod_{2\leq j\leq r}\sin\left[\frac{(M-r+1-k_j)\pi}{N}\right]&\cdots\\
\ldots&\ldots\\\prod_{2\leq j\leq
r}\sin\left[\frac{(M-k_j)\pi}{N}\right]&\cdots \end{array}\right],
\end{eqnarray*}
so the rank of $A^{(r(1))}=r=C_r^1$.

Next consider $A^{r(2)}$, the element is simplified as $\prod_{3\leq
j\leq r}\sin[(k_1-k_j)\pi/N]\sin[(k_2-k_j)\pi/N]$, which is a
product of $2(r-2)$ sine functions. When we simplify $A^{r(2)}$ by
similar strategy, we must bear in mind that both $k_1$ and $k_2$
vary in the value range $[1,M]$ and $k_1<k_2$ must be satisfied. So
the simplest form of $A^{r(2)}$ is
\begin{eqnarray*}
 A^{r(2)}=\left[\begin{array}{ccc}\cdots &\prod_{3\leq j\leq r}\sin\left[\frac{(M-r+1-k_j)\pi}{N}\right]\sin\left[\frac{(M-r+2-k_j)\pi}{N}\right]&\cdots\\
\vdots &\vdots&\vdots\\\cdots &\prod_{3\leq j\leq r}\sin\left[\frac{(M-r+1-k_j)\pi}{N}\right]\sin\left[\frac{(M-k_j)\pi}{N}\right]&\cdots\\
\vdots &\vdots&\vdots\\\cdots &\prod_{3\leq j\leq
r}\sin\left[\frac{(M-1-k_j)\pi}{N}\right]\sin\left[\frac{(M-k_j)\pi}{N}\right]&\cdots
\end{array}\right],
\end{eqnarray*}
the rank of $A^{r(2)}=1+2+\cdots+r-1=C_r^2$.

Finally, analogous but involved generalization yields the ultimate
expression of $A^{r(l)}$ is
\begin{eqnarray*}
 A^{r(l)}=\left[\begin{array}{ccc}\cdots&\prod_{l+1\leq j\leq r}\sin\left[\frac{(M-r+1-k_j)\pi}{N}\right]
\cdots\sin\left[\frac{(M-r+l-k_j)\pi}{N}\right] &\cdots\\
\vdots&\vdots&\vdots\\\cdots&\prod_{l+1\leq j\leq
r}\sin\left[\frac{(M-l+1-k_j)\pi}{N}\right]\cdots
\sin\left[\frac{(M-k_j)\pi}{N}\right]&\cdots \end{array}\right]
\end{eqnarray*}
and its rank is
\begin{eqnarray*}
&&(r-l+1)+(r-l)(1+2)+(r-l-1)(1+2+3)\\&&+\cdots+(1+2+\cdots r-l+1)\\
&=&\sum_{j=1}^{r-l+1}[r-(l-2+j)]\left[\sum_{i=1}^ji\right]\\&=&C_r^l.
\end{eqnarray*}

So far we have explained Eq. (\ref{7}), accordingly Eq.(\ref{8}) is
proved for arbitrary $r$ and $N$. So we have proved that in
arbitrary $N$-qubit $XX$ model, no matter $N$ is infinite or finite,
the occurrence of all QPT, first-order and continuous, can be
witnessed by the change of ground-state entanglement class.

\section{Conclusion}
In conclusion, we have studied the relation between classification
of ground-state entanglement and QPT in $N$-qubit $XX$ model. For
arbitrary $N$, when the exchange constant $J$ is kept invariable and
the external magnetic field $B$ is tuned, QPT occurs. For
$N\rightarrow \infty $ or limited $N$, the QPT is continuous or
first-order respectively. No matter what the type of QPT is, we find
that the \emph{occurrence of QPT is always indicated by the change
of class of entangled ground state}. Although the conclusion
obtained in the $XX$ model seems too particular, we think it is
indeed a reasonable conclusion that the entangled ground states in
the vicinity of the transition are SLOCC-inequivalent. Because the
intrinsic feature of QPT is the change of the structure of the
ground state, and inequivalently entangled states have different
structure. We believe our results grasp the essence of the relevance
of entanglement in QPT and hope it can be verified broadly in the
future. Besides, the proof based on the Schmidt rank provides a
partial solution to judge SLOCC-inequivalent entanglement. Although
it works only if Schmidt rank indeed changes at the QPT, since the
coincidence of the Schmidt rank does not ensure the same
entanglement class for three or more components, it develops a new
method regardlessly the complete classification of multipartite
entangled is far to be resolved nowadays.

\ack The project is supported by National Natural Science Foundation
of China under Grant No. 11174370, 11004248 and 11074307. Chen
thanks Prof. Wei-Ping Zhang and Shi-Yao Zhu for their helpful
discussion.

\end{document}